\documentclass[twocolumn]{raa}
\usepackage{graphicx,times}
\usepackage{natbib}
\usepackage{amssymb,amsmath}
\bibpunct{(}{)}{;}{a}{}{,}

\newcommand{\kms}{~\rm kms^{-1}}

\newcommand{\Msunh}{{~\rm h^{-1}~M_\odot}}

\newcommand{\mpc}{~\rm Mpc}

\newcommand{\mpch}{~\rm h^{-1}Mpc}

\newcommand{\reffig}[1]{Fig.~\ref{#1}}

\usepackage[a4paper=true,driverfallback=dvipdfm,pagebackref=true]{hyperref}

\hypersetup{colorlinks = true, linkcolor = green, anchorcolor = red, citecolor = blue, filecolor = red, pagecolor = red, urlcolor = red}

\begin{document}
 \title{Subhalo abundance and satellite spatial distribution in Milky Way-Andromeda-like paired haloes}

 \volnopage{ {\bf 20XX} Vol.\ {\bf X} No. {\bf XX}, 000--000}
   \setcounter{page}{1}

\author{Kemeng Li \inst{1,2}, 
   Shi Shao \inst{2},
   Ping He \inst{1,3} \thanks{E-mail: hep@jlu.edu.cn},
   Qing Gu \inst{2,4},
   Jie Wang \inst{2,4}
   }


    \institute{ College of Physics, Jilin University, Changchun 130012, China.\\
    \and
             National Astronomical Observatories, Chinese Academy of Sciences, Beijing 100101, China.\\
	\and
    Center for High Energy Physics, Peking University, Beijing 100871, China. \\
    \and
    School of Astronomy and Space Science, University of Chinese Academy of Sciences, Beijing 100049, China\\
 \vs \no
   {\small Received 20XX Month Day; accepted 20XX Month Day}
}

\abstract{We study the subhalo and satellite populations in haloes similar to the Milky Way (MW)-Andromeda paired configuration in the Millennium II and P-Millennium simulations. We find subhaloes are $5\%-15\%$ more abundant in paired haloes than their isolated counterparts that have the same halo mass and large-scale environmental density. Paired haloes tend to reside in a more isotropic environment than isolated haloes, the  shear tensor of their large-scale tidal field is possibly responsible for this difference. We also study the thickness of the spatial distribution of the top 11 most  massive satellite galaxies obtained in the semi-analytic galaxy sample constructed from the Millennium II simulation. Moreover, satellites that have lost their host subhaloes due to the resolution limit of the simulation have been taken into account. As a result, we find that the difference in the distribution of the satellite thickness between isolated and paired haloes is indistinguishable, which suggests that the paired configuration is not responsible for the observed plane of satellites in the Milky Way. The results in this study indicate the paired configuration could bring some nonnegligible effect on the subhalo abundance in the investigation of the Milky Way's satellite problems. 
}

   \authorrunning{Li et al.}            
   \titlerunning{Subhalo \& Satellites in MW-M31-like paired haloes}  
   \maketitle
%
\section{Introduction}
\label{sec:intro}  
The $\Lambda$ cold dark matter ($\Lambda$CDM) paradigm explains successfully the large-scale structure of our Universe, including the anisotropies of the cosmic microwave background (CMB) radiation or galaxy clustering on large scales. However, the observed satellites in our Local Group (LG), as the primary probes of small-scale structures, pose challenges to our cosmological model \citep[e.g.][]{mateo1998dwarf, bullock2017small}. The most controversial debates include, for example, the `missing satellites' problem, i.e., the overabundance of predicted dark matter substructures compared to observed luminous satellites in our Milky Way (MW) \citep{moore1999dark}, and the `too-big-to-fail' (TBTF) problem, i.e., that the predicted most massive subhaloes in simulations are too dense to host the observed most luminous satellites \citep{boylan2011too, boylan2012milky}.

Many studies investigated how interactions between the host halo and its satellites can reconcile the $\Lambda$CDM paradigm with the puzzles by introducing baryonic physics, e.g. reionization, SN feedback, and environmental effects, e.g. stellar stripping and tidal destruction \citep[e.g.][]{diemand2007formation, maccio2010luminosity, boylan2012milky, garrison2013can, brooks2014baryons, dutton2016nihao}. However, these studies have yet failed to solve the small-scale problems, and more works attempt to resolve this problem from alternative perspectives. \cite{wang2012missing} suggests that the lack of massive subhaloes simply indicates an overestimate of the mass of the MW. \cite{cautun2014subhalo} tried to extrapolate subhalo abundance beyond the mass resolution limit of $N$-body simulations and found $\sim$20\% more substructures in MW-mass haloes than those in other studies. \cite{brook2015expanded} considers those dwarf galaxies with $M_{*}>10^{6}\ \mathrm{M_{\odot}}$ can be assigned to more massive haloes with mass-dependent halo profiles. \cite{sawala2016apostle} found a good agreement between the observed satellite population in the LG and those in the simulated paired haloes with similar kinematical configuration to our LG by adjusting the physical models such as reionization and SN feedback in $\Lambda$CDM simulations. \cite{smercina2018lonely} suggests that satellite galaxies perhaps form much more stochastically than the dramatically increasing slope predicted by the `standard' halo occupation models \citep[e.g.][]{behroozi2013average}.

Besides the `missing satellites' and the 'TBTF' problems, another puzzling feature of our LG is that the classical satellites of the MW lie on a spatially thin plane \citep[e.g.][]{Lynden-Bell_76, kroupa2005great,Shao_16}, and with many of the satellites being orbiting within the plane \citep[e.g.][]{Metz2008, Pawlowski2013b, shao2019evolution, Fritz2018}. \cite{cautun2015planes} addressed that $\sim 10\%$ of $\Lambda$CDM haloes have spatially thin satellite planes and they are even more prominent than those present in the Local Group. Nevertheless, \cite{shao2019evolution} showed that the chance of having a kinematically thin satellite plane in $\Lambda$CDM simulations is less than $1\%$, and the satellite plane can be long-lived and can persist for at least 4 Gyrs. \citet{samuel2020planes} also argued that for hosts with an LMC-like satellite near the first pericenter, the longevity of MW-like planes could even increase $\sim$3 Gyr.

Several studies \citep{libeskind2005distribution, libeskind2011preferred} showed that the planes of satellites are a consequence of highly anisotropic accretion such as the torques exerted by the surrounding large-scale structures, like an M31, which tilt the satellite orbits onto the plane. This is most important for this work, we therefore proceed by identifying in the N-body cosmological simulations paired halo systems similar to the MW-M31 configuration. We explore whether the existence of a massive companion can affect the subhalo abundance of an MW-like halo. Also, we will study the abundance and spatial distribution of the satellite galaxies generated by the semi-analytic galaxy formation model. The paper is organized as follows. In Section~\ref{sec:simsam}, we present the details of the simulations and the sample selection criteria. In Section~\ref{sec:results}, we present the main results. We present our conclusions and discussions in Section~\ref{sec:condis}.

\section{Simulation and Samples}
\label{sec:simsam}

\subsection{Simulations}

We make use of two sets of $N$-body cosmological simulations: Millennium Simulation-II (MS-II, \cite{boylan2009resolving}) and Planck-Millennium simulation (P-Mill, \cite{jiang2014, baugh2018galaxy}). MS-II traces the evolution of $2160^{3}$ dark matter particles from 127 to 0 in a periodic box with a side length of 137 \mpc. The dark matter particle mass is $9.45 \times 10^{6}\ M_{\odot}$. MS-II uses cosmological parameters as follows: $\Omega_{\mathrm{m}} = 0.25$, $\Omega_{\Lambda} = 0.75$, $\Omega_{\mathrm{b}} = 0.045$, $\sigma_{8} = 0.9$ and $H_{0} = 73 \mathrm{km\ s^{-1} Mpc^{-1}}$. P-Mill has a larger box than MS-II with a side length of 800 Mpc. The simulation has $5040^{3}$ dark matter particles with a mass of $1.565 \times 10^{8}\ M_{\odot}$. P-Mill adopts a \textit{Planck} cosmology with parameters: $\Omega_{\mathrm{m}} = 0.307$, $\Omega_{\mathrm{\Lambda}} = 0.693$, $\sigma_{8} = 0.8288$ and $H_{0} = 67.77\ \mathrm{km\ s^{-1} Mpc^{-1}}$.

In order to study the satellite population in the simulation, we make use of a semi-analytic model provided by \citet{guo2011dwarf} based on the MS-II. Semi-analytic model of galaxy formation is flexible and computationally efficient in generating synthetic galaxy populations. The philosophy adopted in the model is based on the (sub)haloes in merger trees. Galaxies are populated into (sub)haloes by solving a set of differential equations that govern the cooling of gas in haloes, star formation, feedback from stars and black holes, chemical enrichment, and the evolution of stellar populations. The free parameters in the model are calibrated according to a selection of observed properties of the local galaxy population. We refer the reader to \citet{guo2011dwarf} for more detailed descriptions.

In both simulations, haloes were identified using the friends-of-friends (FOF) method \citep{davis1985evolution}. The linking length is 0.2 times the mean particle separation. The haloes were then processed to identify gravitationally bound substructures performed by applying the SUBFIND code \citep{springel2001populating} to the dark matter particles associated with each FOF halo. The objects were split into main haloes and subhaloes. The main haloes are characterized by their mass, $M_{\rm 200}$, and radius, $R_{\rm 200}$, which correspond to an enclosed spherical overdensity of 200 times the critical density.

\subsection{Sample selection}

\begin{figure*}
\centering
\includegraphics[width=2\columnwidth]{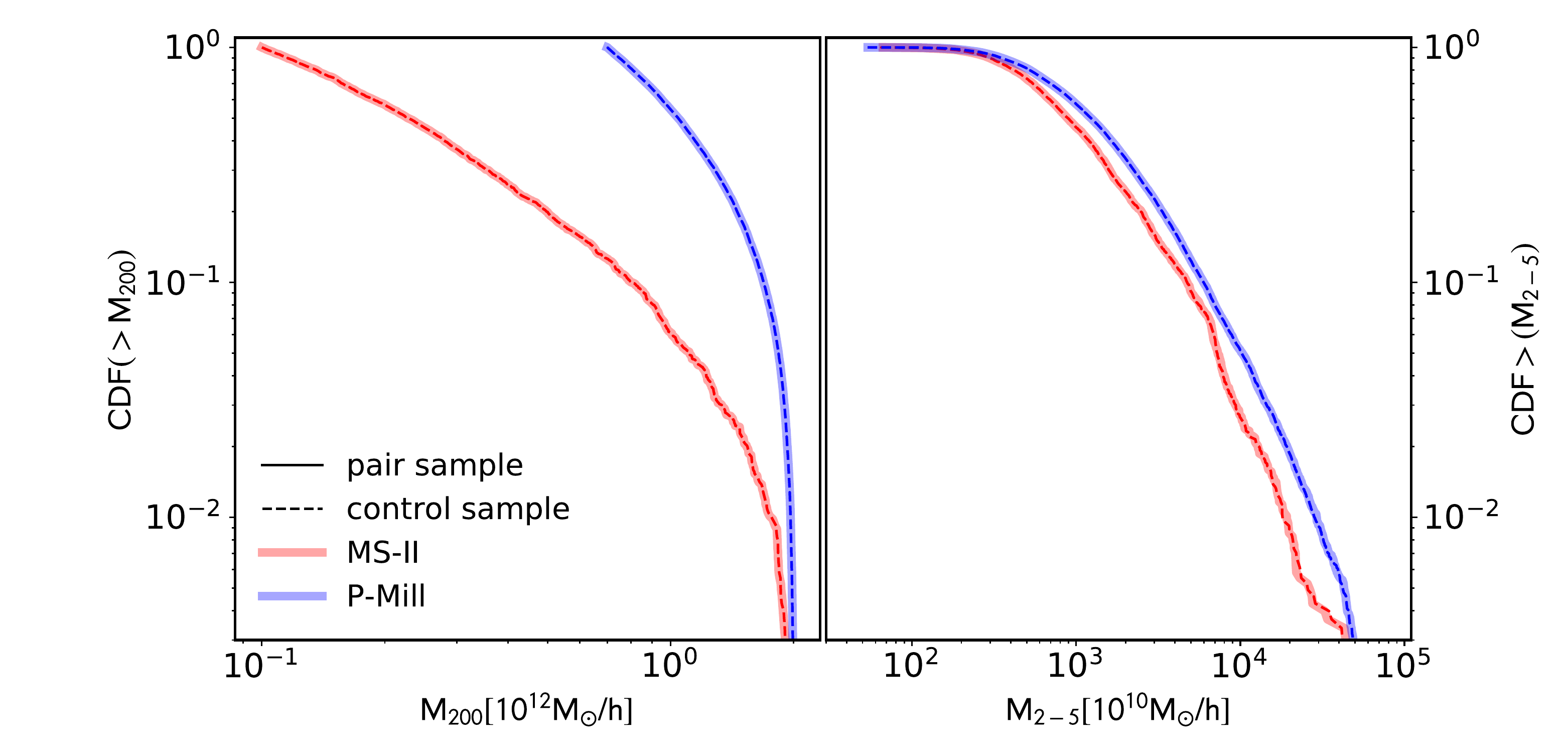}
\caption{\textit{Left panel}: The cumulative distribution functions of $M_{\mathrm{200}}$ of hosts in pair samples (solid lines) and control samples (dashed lines) in P-Mill (red lines) and MS-II (blue lines). \textit{Right panel}: The cumulative distribution functions of the environmental density, $\mathrm{M_{\rm 2-5}}$ of hosts in the two samples.} 
\label{M200} 
\end{figure*} 

\begin{figure}
\centering
\includegraphics[width=\columnwidth]{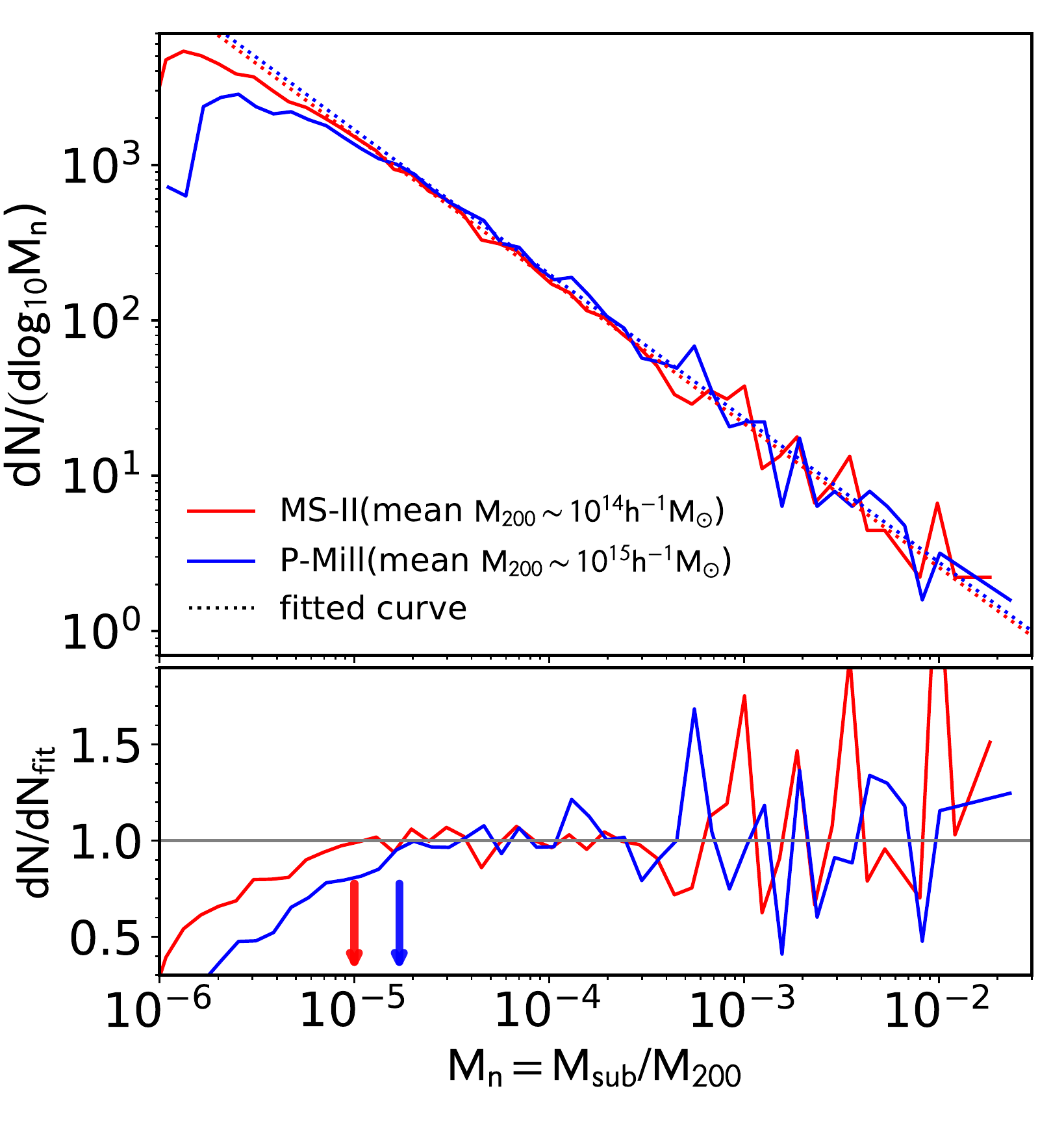}
\caption{\textit{Top panel}: the differential subhalo abundance as a function of the scaled subhalo mass $M_{\mathrm{n}} \equiv M_{\mathrm{sub}}/ M_{\mathrm{200}}$ of clusters in P-Mill (solid blue, $\alpha \sim 0.927$) and MS-II (solid red, $\alpha \sim 0.925$). The mean $M_{\mathrm{200}}$ of the five clusters is $\sim 1.0 \times 10^{15} h^{-1}\mathrm{M_{\odot}}$ and $1.4 \times 10^{14} h^{-1} \mathrm{M_{\odot}}$ in P-Mill and MS-II, respectively. The dashed lines represent the fitted mass functions. \textit{Bottom panel}: the ratio of the differential subhalo abundance to the fitted value as a function of $M_{\mathrm{n}}$ in MS-II and P-Mill. The arrows represent the smallest subhalo mass $1.4 \times 10^{9} h^{-1}\mathrm{M_{\odot}}$ and $1.7 \times 10^{10} h^{-1}\mathrm{M_{\odot}}$ for MS-II and P-Mill, respectively, i.e. $N_{\rm crit} \sim 200$ in both simulations.} 
\label{T_err} 
\end{figure}

We want to work with a representative sample of similar halo mass to our MW. Thus, we select MW-mass haloes by requiring the host halo mass to be in the range $M_{\rm 200} \in [0.1, 2]\times10^{12}\Msunh$ for the MS-II simulation, and a smaller mass range $M_{\rm 200} \in [0.7, 2]\times 10^{12} \Msunh$ for the P-Mill simulation since its mass resolution is roughly 10 times lower than MS-II. The wide mass range is due to the large uncertainties in the estimates of MW halo mass \citep[e.g.][]{callingham2019mass,wang2020mass}. We then select the Local Group-like sample by requiring that two MW-mass haloes lie within a separation distance in the range $d\in[0.7, 1.2]\mpc$. We require that any such paired haloes not overlap with any other more massive haloes. Thus, we exclude all paired haloes that have a companion within $1.4\mpc$ with its mass larger than any of the paired haloes. We refer to the haloes in the Local Group-like sample as the \textit{pair} sample.

Generally speaking, paired haloes are formed in high-density regions, such that they have more subhaloes than their isolated counterparts with the same halo mass \citep[e.g.][]{ishiyama2008environmental}. For a fair comparison between the two samples, we require the difference in environment density ($\mathrm{log_{10}(M_{2-5})}$) between the two samples to be smaller than 0.01. $\mathrm{M_{2-5}}$ is defined as the total mass of dark matter particles within a spherical shell in the range of 2-5Mpc. As studied in \citet{Han18}, the DM density of outer regions of DM haloes is a good proxy for modeling the large-scale environment of galaxies. Since more massive haloes tend to have more subhaloes \citep{gao2004subhalo}, we further select haloes from the MW-mass sample to have a counterpart in the pair sample with the minimum difference in their halo mass being smaller than 0.01. We refer to these MW-mass haloes as the \textit{control} sample.

In this work, we take into account the subhaloes that reside within $R_{\rm 200}$ from the centre of each halo. To eliminate the affection caused by the accretion of extremely massive subhaloes \citep[e.g.][]{Wang2013,Deason2014,Shao2018b,Shao2018a}, we further exclude haloes from the sample if any of its subhaloes satisfies $M_{\rm sub}/M_{\rm host}>0.15$, where $M_{\rm sub}$ is the sum of the mass of the bound DM particles that are associated with this subhalo. Finally, we obtain 3263 and 18871 control/pair samples that satisfy our selection criteria in MS-II and P-Mill, respectively. The distributions of the $M_{\rm 200}$ and $M_{2-5}$ of these samples are shown in \reffig{M200}. The control and pair samples have almost identical distributions in either of the simulations. This indicates that the effect on the satellite abundance caused by the background density has been significantly reduced.
 
 As pointed out by \citet{springel2008aquarius}, the subhalo mass function at the low-mass end is strongly affected by the mass resolution of the simulation. Therefore, at least a few hundred particles are required to resolve the smallest subhaloes in low-resolution simulations, to converge with the results in high-resolution simulations. We perform a careful check on this issue in five clusters with mass of $M_{\mathrm{200}} \sim 1.0 \times 10^{15} h^{-1} \mathrm{M_{\odot}}$ and $1.4 \times 10^{14} h^{-1}\mathrm{M_{\odot}}$ in P-Mill and MS-II, respectively. In the top panel of Fig.~\ref{T_err}, we show the differential subhalo mass function and the best-fitting power-law relation of these two samples. The data deviates significantly from the best-fitting line below subhalo mass, $1.4 \times 10^{9} \Msunh$ and $1.7 \times 10^{10} \Msunh$, for MS-II and P-Mill, respectively. The two masses roughly correspond to 200 particles in the two simulations. Thus, we consider subhaloes with at least 200 particles to be completely resolved in the simulations.

\section{Results}
\label{sec:results}

In Section~\ref{sec:ssa}, We present the subhalo and satellite abundance in the pair sample and control sample and discuss the causes for the difference in the two samples. In Section~\ref{sec:ssd}, we present the spatial distribution of the 11 most massive satellites in the two samples, which includes satellite radial distribution and the thickness of the plane of satellites.

\begin{figure*}
\centering
\includegraphics[width=2\columnwidth]{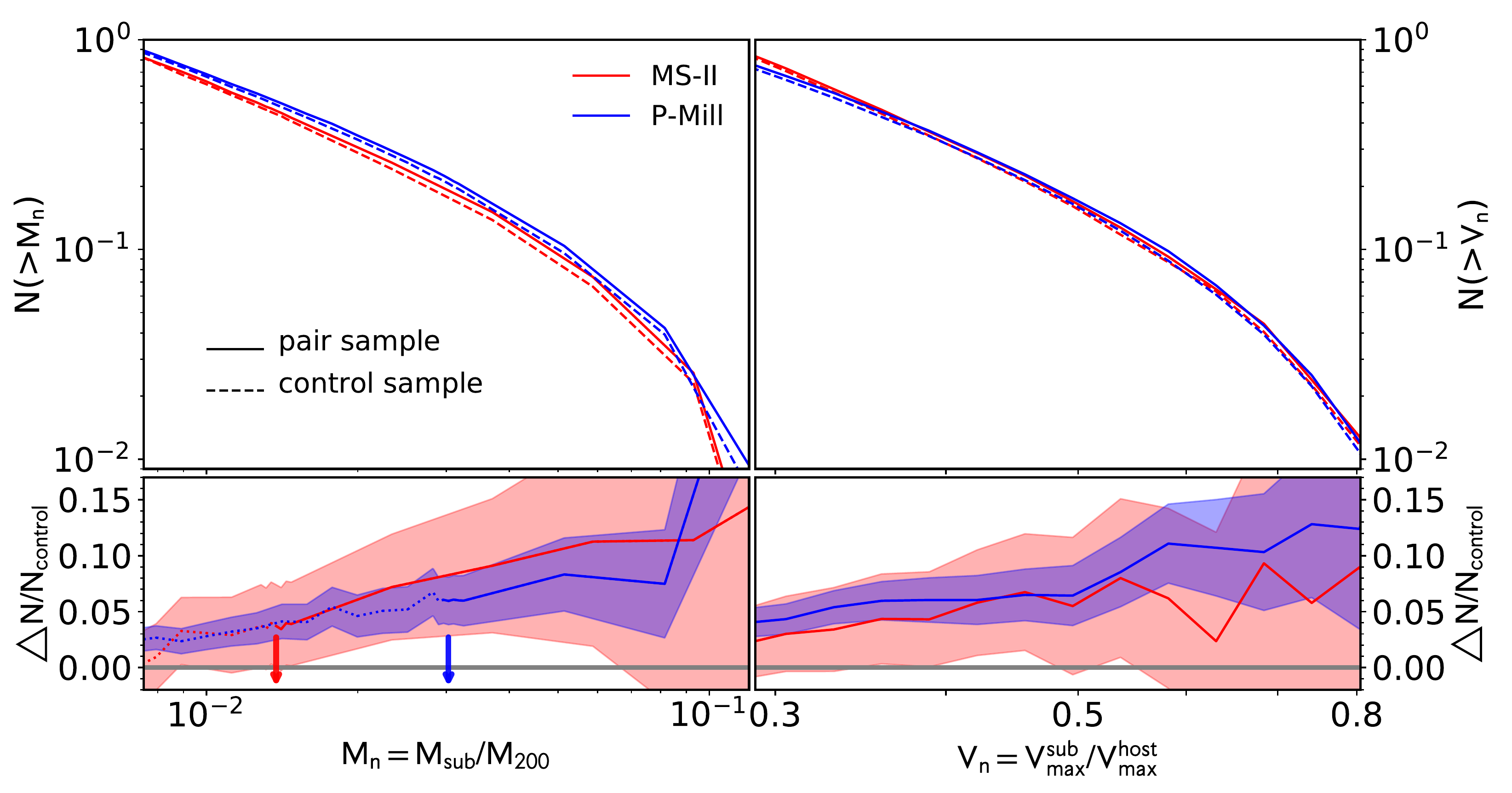}
\caption{The subhalo abundance as a function of the scaled subhalo mass $M_{\mathrm{n}} \equiv M_{\mathrm{sub}}/ M_{\mathrm{200}}$ (\textit{left panel}) and the scaled maximum circular velocity $\mathrm{V_{n} \equiv V_{max}^{sub}/V_{max}^{host}}$ (\textit{right panel}) in pair (solid lines) and control samples (dashed lines) in MS-II (red lines) and P-Mill (blue lines). Bottom left: The relative difference of well-resolved subhalo abundance as a function of $M_{\mathrm{n}}$ in MS-II (red solid line) and P-Mill (blue solid line). The dotted lines represent subhaloes that are not well-resolved. The arrows indicate the upper limit of critical subhalo masses in both simulations. The shadow regions represent $1 \sigma$ scatter calculated from the \textit{bootstrap} resampling method. Bottom right: The relative difference of the subhalo abundance as a function of $\mathrm{V_{n}}$ in MS-II and P-Mill.}
\label{Nsub} 
\end{figure*}

\begin{figure*}
\centering
\includegraphics[width=2\columnwidth]{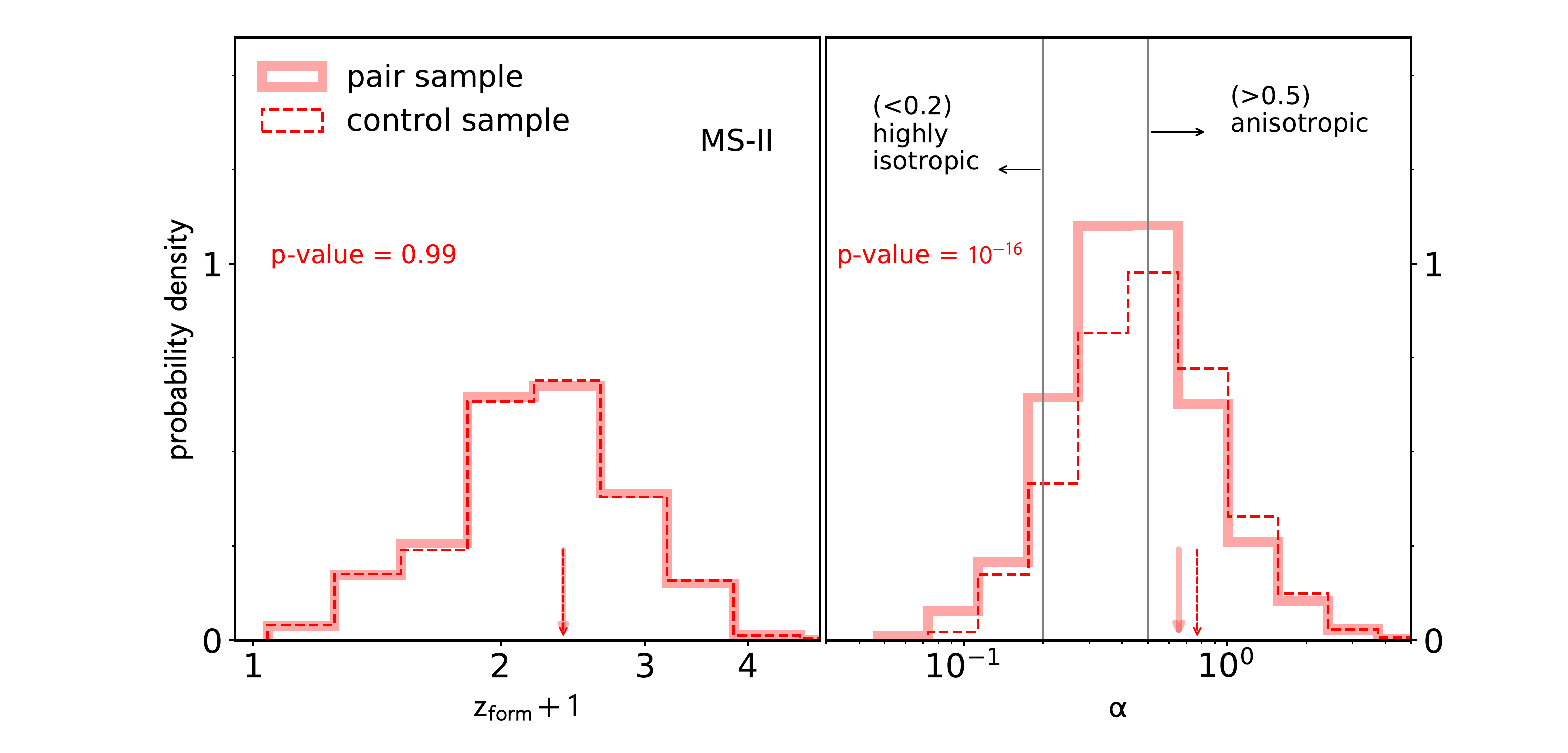}
\caption{\textit{Left panel}: the probability density of the formation redshift of pair sample (red solid) and control sample (red dashed) in MS-II. Vertical arrows represent the median of formation redshift for the two samples which are almost identical, $z\approx1.4$. \textit{Right panel}: the probability density of the tidal anisotropy in halo environments. The median of the distribution is $\alpha=0.65$ for pair sample and $\alpha=0.77$ for control sample. The vertical grey lines represent highly isotropic regime ($\alpha < 0.2$) and anisotropic regime ($\alpha > 0.5$).}
\label{envir} 
\end{figure*}

\subsection{Subhalo and satellite abundance}
\label{sec:ssa}
In this section, we check the abundance of the subhaloes in the pair and control samples in two simulations; we use the \textit{bootstrap} resampling method to estimate the mean value and variance of the subhalo abundance for each sample. We further study their correlation to the formation time and the local environment of host halos. We also study the abundance of satellites in the two samples with the galaxy catalogue extracted from \cite{guo2011dwarf}.

\subsubsection{Subhalo abundance}
\label{sec:suba}

In the top-left panel of Fig.~\ref{Nsub}, we show the cumulative abundance of subhaloes as a function of their scaled subhalo mass $M_{n}$ in the pair and control samples for the two simulations. The arrows indicate the subhalo mass limit due to the resolution effect in the MS-II and the P-Mill. In both simulations, subhaloes in the pair sample are more abundant than those in control samples for all scaled subhalo mass. As shown in the left-bottom panel, the excess of subhalo abundance ($\delta N/N=N_{\mathrm{sub, pair}}/N_{\mathrm{sub, control}} - 1$) increases as increasing scaled subhalo mass from $\delta N/N \approx 5\%$ at $M_{\rm n}=0.02$ to $\delta N/N  \approx 12\%$ and $18\%$ at $M_{\rm n}=0.1$ for MS-II and P-Mill, respectively. We have to note that the subhalo overabundance is mainly contributed by the more massive haloes in the pair sample, which is shown in Appendix \ref{sec:M1M2}.

We have to note that the difference in the subhalo abundance between MS-II and P-Mill is non-negligible. This could be caused by the uncertainty in the estimate of subhalo mass. Therefore, we use an alternative quantity, the maximum circular velocity $V_{\mathrm{max}}$, to cross-check the result. We define the scaled maximum circular velocity $V_{\rm n}$ as $V_{\rm max}^{\rm sub}/V_{\rm max}^{\rm host}$, where $V_{\rm max}^{\rm host}$ is the circular velocity of the host halo. The result is shown in the right panel of \reffig{Nsub}. We find a good agreement between the two simulations and an excess of subhalo abundance in the pair sample compared to control samples as a function of the scaled maximum circular velocity $V_{\rm n}$. Particularly, in P-Mill, the subhaloes are 10\% more abundant in the pair sample than those in the control sample for subhaloes with $V_{\rm n} > 0.6$ which is equivalent to $V_{\rm max}$ $\sim 60 \kms$; this is the value for the most massive satellite, LMC, in our MW. This indicates that the change of having an LMC-mass satellite is enhanced in paired haloes like our Local Group.

\subsubsection{Formation time and environmental effect}
\label{sec:envir}
We found subhaloes are more abundant in the pair sample than that in the control sample. We explore the reason for this overabundance in two aspects: 1) the assembly history of the halo; 2) the anisotropy of the halo's cosmic web environment. We first check the halo assembly history via its formation redshift $z_{f}$ that is the time when the halo reaches half of its final mass, $M_{\mathrm{200, z0}}$. The formation redshift of host haloes in MS-II simulation is presented in the left panel of Fig.~\ref{envir}. We find that the distributions of halo formation time in the control sample and pair sample are almost identical, which indicates that the entire halo assembly history is not responsible for the difference in the subhalo abundance. Nevertheless, we have to note that subhaloes at infall can experience strong tidal stripping by their host halo \citep{Han16}, this stripping process could be very different between the two samples.

We exclude the effect on halo properties by environment density when constructing our samples. Here, we quantify the effect of another environmental factor, $\alpha$, the tidal anisotropy of a halo's cosmic web environment, on the abundance of subhaloes. The parameter, $\alpha$, is defined as $\sqrt{q^{2}/(1+\delta)}$, where $q$ is the halo-centric tidal shear \citep{heavens1988tidal,catelan1996evolution}, $\delta$ is the halo-centric overdensity. For more details, we refer the reader to \citet{paranjape2018halo} (see also \cite{ramakrishnan2019cosmic}). By definition, haloes that reside in highly isotropic local environments have a low value of $\alpha <0.2$ while haloes that reside in the anisotropic environment (e.g. filament) have a large $\alpha >0.5$. As shown in the right panel of Fig.~\ref{envir}, the tidal anisotropy of pair and control samples show significantly different distributions with a KS test p-value of $10^{-11}$. We find that there  are 2.62\% and 1.68\% of haloes reside in highly isotropic environments ($\alpha < 0.2$ ), while there are 66.69\% and 75.14\% of haloes reside in anisotropic, filament-like environments ($\alpha > 0.5 $) for pair and control samples, respectively. This result suggests that it is more likely to find a paired halo in the isotropic environment than that in the anisotropic environment.

To confirm this result, we also classify the environment into 4 different types: void, sheet, filament, and knot according to the tidal tensor around the halo \citep{forero2009dynamical}. This classification is based on the evaluation of the Hessian of the gravitational potential. We use a smoothing scale, $R_{s} = 0.8 \mpch$, and a threshold eigenvalue, $\lambda_{th}= 0.1$, for the calculation. We find that 13.78\% of the pair sample reside in knots that correspond to the isotropic environment; this fraction is nearly 50\% higher than the control samples with only 9.26\% of the sample residing in knots. This agrees with the result obtained from the calculation of tidal anisotropy. Thus, the overabundance of subhalo in the pair sample could be enhanced in the isotropic, knot-like environment. This is consistent with the previous study \citep{metuki2015galaxy}.

\subsubsection{Satellite abundance}
\label{sec:sata}
\begin{figure}
\centering
\includegraphics[width=\columnwidth]{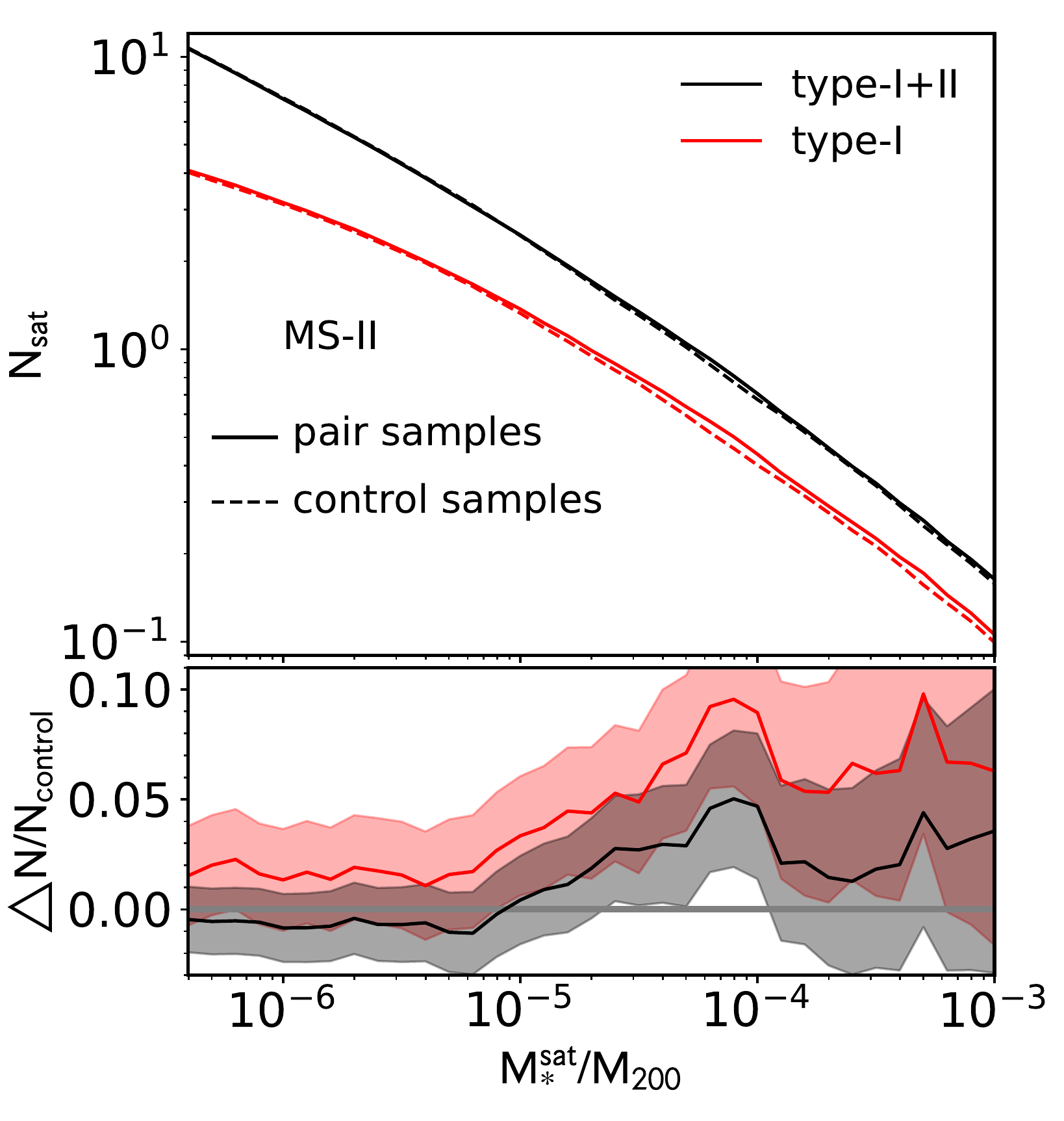}
\caption{\textit{Top panel}: satellite abundance as a function of scaled stellar mass, $M^{\rm sat}_{\star}/M_{\rm 200}$ in pair sample (solid lines) and control sample (dashed lines) for type-I satellites (red lines), and for type-I+II satellites (black lines) in MS-II. \textit{Bottom panel}: relative difference of satellite abundance of type-I (red line) and type-I+II satellites (black line) between pair and control samples. The shadow regions represent 1$\sigma$ scatter calculated from the \textit{bootstrap} resampling method.}
\label{Nminus_sat} 
\end{figure}

From the observational perspective, it is also interesting to study the luminous satellite abundance in the two samples. We therefore evaluate the difference in satellite abundance as a function of stellar mass $M^{\mathrm{sat}}_{\star}$ between the pair sample and control sample by making use of the semi-analytic galaxy catalogue \citep{guo2011dwarf}. In the semi-analytic model, we can keep track of ``orphan satellite galaxies'', whose subhaloes have been disrupted below SUBFIND's detection limit of 20 particles after falling into the host halo. We refer to these ``orphan satellite galaxies'' as Type-II satellites and the rest with surviving subhaloes as Type-I satellites. 

As shown in Fig.~\ref{Nminus_sat}, we find Type-I satellites are at least $\sim5\%$ more abundant in the pair sample than that in the control sample, which is in accordance with the result of subhalo mass distributions in MS-II. However, when Type-II satellites are taken into account, pair sample only shows a very weak excess ($\sim2\%$) in the satellite abundance at the high-mass end, albeit with a very large scatter. This suggests that the large discrepancy in the subhalo abundance is mainly due to the fact that massive subhaloes have experienced very different tidal stripping processes between the control and pair samples. We further check this result by tracking back all the satellites prior to their infall when their subhalo mass peaked, and find that the subhaloes in the pair sample are also $\sim2\%$ more abundant than those in the control sample, which is consistent with the result for their satellites at $z=0$ (see Appendix~\ref{sec:Mpeak}.)

\subsection{Satellite's spatial distribution}
\label{sec:ssd} 
Most of the MW satellites define a tight plane \citep[e.g.][]{Kunkel_76, kroupa2005great, Lynden-Bell_76, Shao_16} that shows some degree of coherent rotation \citep[e.g.][]{pawlowski2013dwarf, shao2019evolution}. In this section, we investigate whether a massive companion halo, like M31, could affect the spatial distribution of the 11 most massive satellites in the MW. The satellites are ranked by their stellar mass according to the results of the semi-analytic model. Because of the mass resolution limit, even in MS-II, the subhalo of the top 11 satellites cannot be fully resolved. For each of these satellites, namely Type-II galaxies, a position is assigned by tracking the most bound particle of the host subhalo from the time it was last resolved. As \citet{wang2012spatial} pointed out, the radial distribution of the top 11 galaxies including these galaxies is in reasonable agreement with their higher resolution counterparts. So here we include these type-II galaxies in our sample.

\begin{figure}
\centering
\includegraphics[width=\columnwidth]{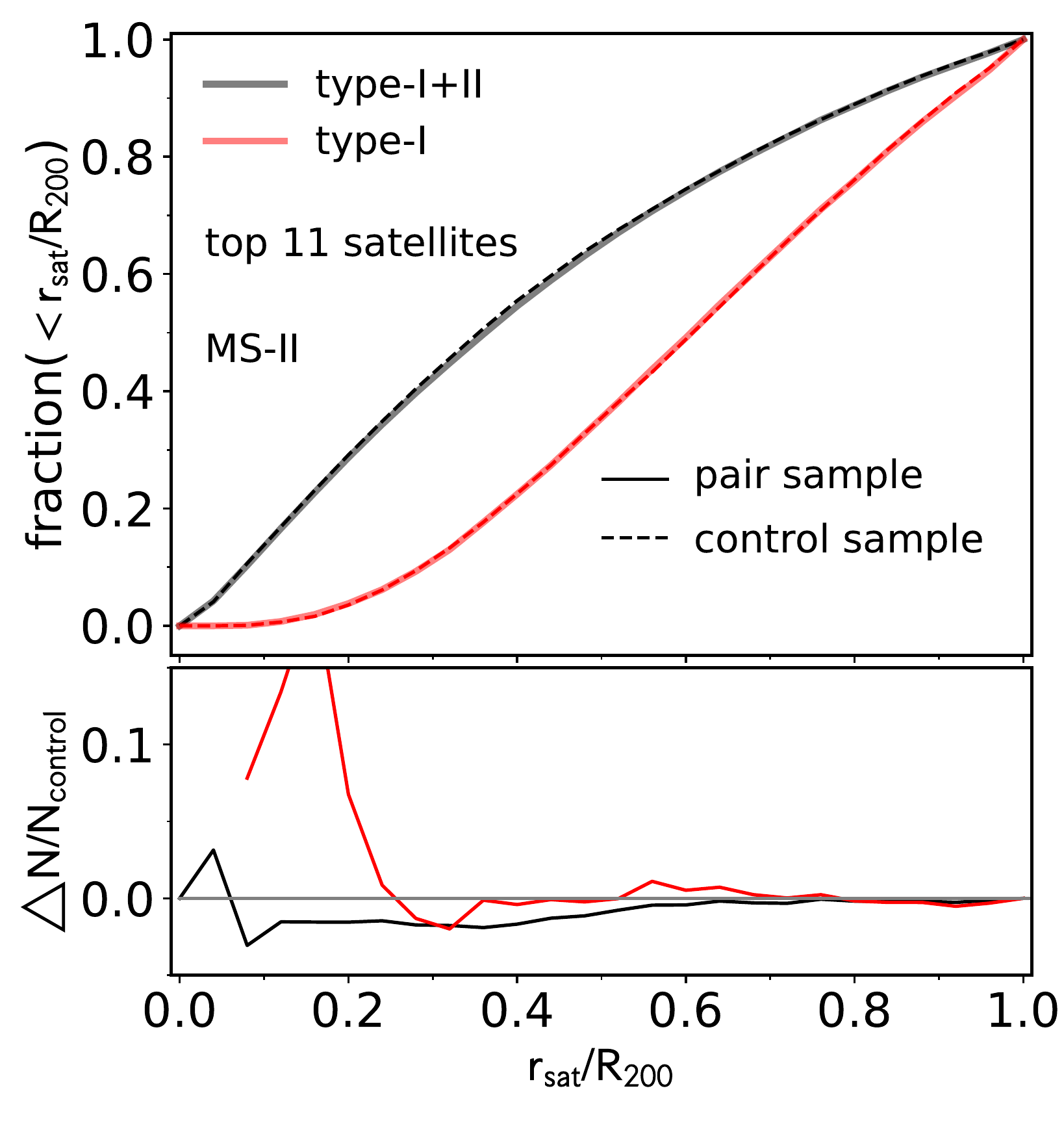}
\caption{\textit{Top panel}: the cumulative radial distribution for the top 11 satellites within $R_{\mathrm{200}}$ from the centre of the hosts in pair sample (solid line) and control sample (dashed line). The black lines represent all the top 11 satellites; the red lines represent only the type-I in the top 11 satellites. \textit{Bottom panel}: The ratio of satellite fraction in the pair sample to that in the control sample.}
\label{radial}   
\end{figure}

\begin{figure}
\centering
\includegraphics[width=\columnwidth]{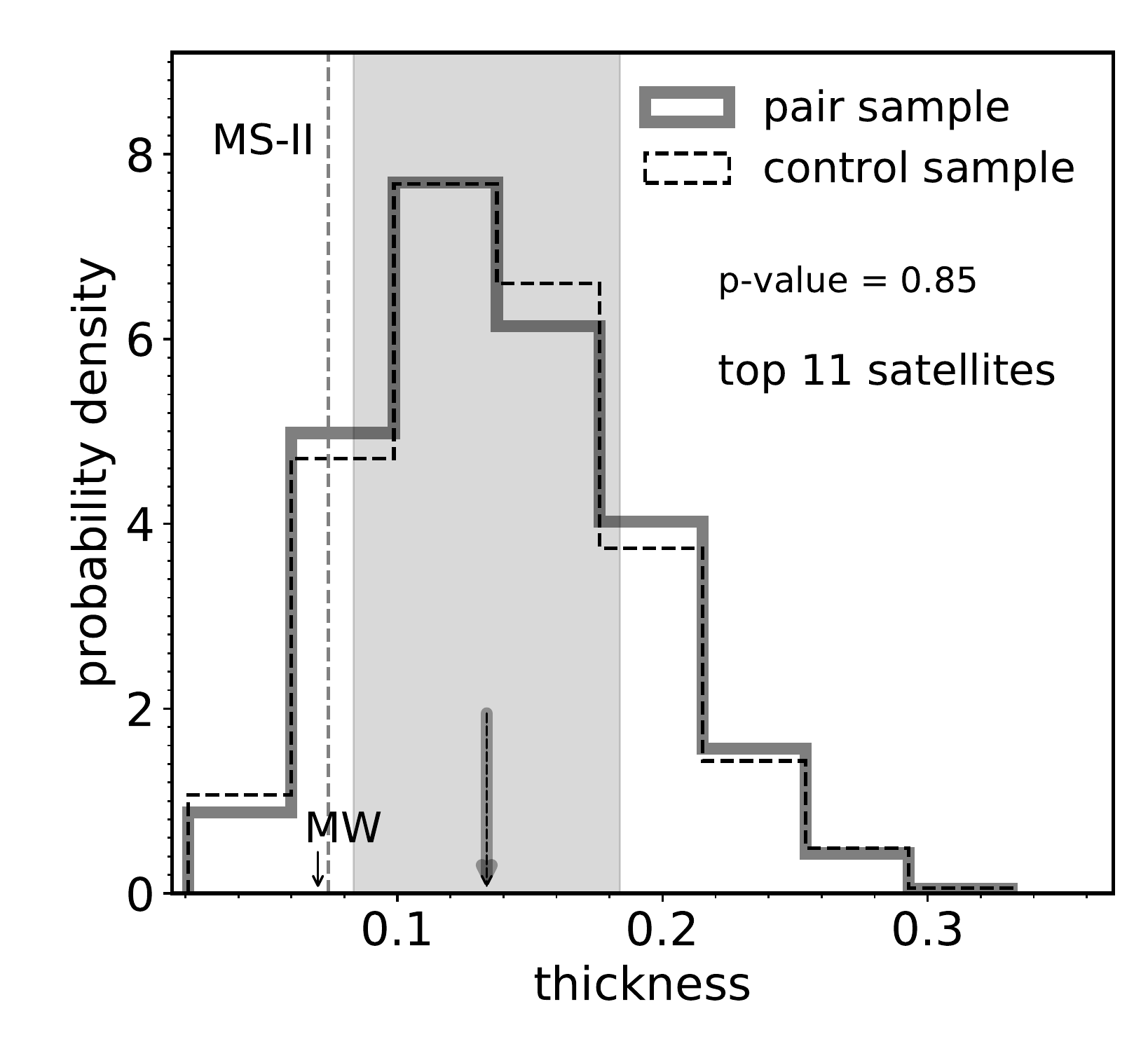}
\caption{The probability density of thickness of the planes of the top 11 satellites in the pair sample (solid black) and in the control sample (dashed black). The two distributions have almost identical median values of $\sim$ 0.14 (black arrows), and the standard error of $\sim$ 0.05 (grey shaded region). The vertical dashed grey line represents the thickness of the measurement of the MW, 0.074.}
\label{Thn}   
\end{figure}

\subsubsection{Radial distribution of the top 11 satellites}
\label{sec:radis}  
We present the cumulative radial distribution of the top 11 satellites in the pair and control sample in Fig.~\ref{radial}. We stacked the satellite radial distances normalised by the $R_{\rm 200}$ of their host haloes. Type-I satellites are more abundant at the inner halo in the pair sample than those in the control sample. However, when Type-II satellites are accounted for, the difference between the two samples becomes negligible. This suggests that tidal stripping is more efficient in the pair sample than in the control sample.

\subsubsection{The thickness of the satellite plane}
\label{sec:thickness} 
Here, we quantify the thinness of the top 11 satellites for each system by fitting the positions of satellites to a best-fitting plane with the RMS of the heights of satellites to the plane being the minimum. Following the definition by \cite{kang2005great} and \cite{kroupa2005great}, thickness is defined as $h_{\rm rms}/R_{\rm cut}$, where $R_{\rm cut}$ is the radial extent of the satellite system which is equivalent to $R_{\rm 200}$ in this study. In Fig.~\ref{Thn}, we present the distribution of the thickness of the satellite planes for the pair and control sample. The thickness of the 11 classical satellites of the MW is 0.074, which is at the tail of the distribution from the MS-II simulation, and only $\sim7\%$ of the simulated satellite systems are even more prominent than the observed value of the MW's satellites, which agrees with the results in previous literatures \citep[e.g.][]{wang2012spatial,Shao_16}. The two distributions have the same standard variance of 0.05 which is shown in the shaded region. The difference between the two distributions is indistinguishable with a K-S test p-value of 0.85.  

\section{Conclusion and Discussion}
\label{sec:condis}

In this work, we use two simulations, P-Mill and MS-II to study the subhalo abundance in Local Group-like (pair) and MW-like (control) haloes. We study the dependence of subhalo abundance on the assembly history of the host halo and the surrounding large-scale structure. We also explore the satellite abundance and the flatness of satellite planes in pair and control haloes with the semi-analytic galaxy catalogue extracted from \cite{guo2011dwarf}. Our main results are summarised as follows.

\begin{enumerate}
\item The haloes in the pair sample have $\approx5\%$ more subhaloes than those in the control sample for low-mass subhaloes with scale mass $M_{\rm n}=0.02$. The overabundance increases with increasing subhalo mass with the value being $\approx15\%$ for high-mass subhaloes $M_{\rm n}=0.1$. 

\item Haloes in the two samples have identical formation histories, which suggests that the overabundance is not due to the assembly history of the halo.

\item Haloes in the pair sample tend to reside in an isotropic, knot-like environment. By construction, the two samples have the identical distribution of halo mass and environmental density, thus, the large-scale tidal field should be responsible for the overabundance of subhaloes seen in the pair sample.

\item By using the semi-analytic galaxy catalogue, type-I satellites in the pair sample show a similar subhalo overabundance compared to the control sample.

\item We find the difference in the spatial distribution of the top 11 satellites between the pair and control sample is indistinguishable. This suggests that the plane of satellites seen around our MW is not due to the existence of the M31.

\end{enumerate}
 
Our result is robust for the most massive subhaloes that have deep potential to host the most massive satellites, e.g. LMC, in our MW. However, limited by the resolution of our simulations, the faintest subhaloes and satellites are not completely resolved. Besides, we don't find a significant difference in radial distribution and thickness of the top 11 satellite systems between the pair sample and the control sample. This could be due to the fact that the uncertainties in the positions of those satellites that have lost their subhalo. We expect that a higher-resolution simulation could solve these issues. The origin of the difference in the subhalo mass function perhaps results from the discrepancy in the tidal stripping and evolution of the environment between the pair and control samples, which we will explore in more detail in future studies.


\begin{acknowledgements}
The authors would like to thank Carlos Frenk, and Marius Cautun for valued discussions and suggestions, and we thank the referee for great and detailed comments. This work is supported by the National Key R\&D Program of China (No.2018YFE0202900 and 2022YFA1602901), the NSFC grant (No 11988101, 11873051, 12125302, 12273053). KL, SS, and JW acknowledge support from the K.C.Wong Education Foundation. SS and JW acknowledge the support of CAS Project for Young Scientists in Basic Research, Grant No. YSBR-062. SS acknowledges the science research grants from the China Manned Space Project with NO.CMS-CSST-2021-B03.
PH acknowledges the support of the National Science Foundation of China (No. 12047569, 12147217), and the Natural Science Foundation of Jilin Province, China (No. 20180101228JC). This work used the DiRAC Data Centric system at Durham University, operated by ICC on behalf of the STFC DiRAC HPC Facility (www.dirac.ac.uk). This equipment was funded
by BIS National E-infrastructure capital grant ST/K00042X/1, STFC capital grant ST/H008519/1, and STFC DiRAC Operations grant ST/K003267/1 and Durham University. DiRAC is part of the National E-Infrastructure.
\end{acknowledgements}

\section*{Data availability}
The data used in this paper are available from the corresponding author upon reasonable request.

\bibliographystyle{raa}
\bibliography{a_sat}

\appendix

\section{Subhalo abundance in more massive haloes in the pair sample}
\label{sec:M1M2}
\begin{figure}
\includegraphics[width=235pt]{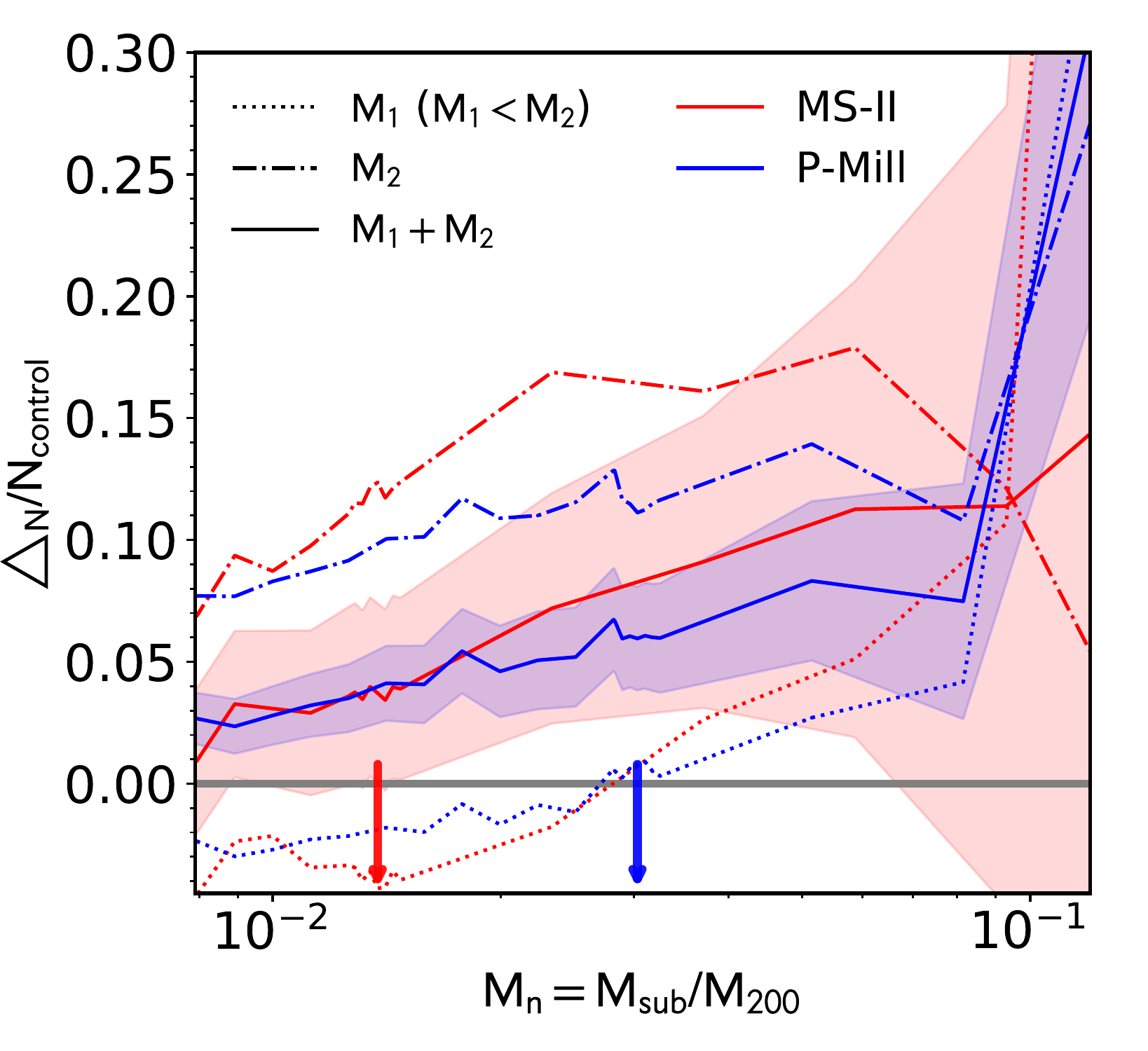}
\caption{The relative difference of subhalo abundance between the pair sample and the control sample in the MS-II (red lines) and P-Mill (blue lines). $M_{1}$ (dotted lines) represents low-mass haloes and $M_{2}$ (dash-dotted lines) for high-mass haloes in the pair sample. The solid line shows the result for the average. The arrows represent the simulation resolution limit (particle number $\geqslant$ 200).} 
\label{Nsub_pair}
\end{figure}

We split the haloes in the pair sample into two subsamples according to the halo mass, $M_{1}$ and $M_{2}$ ($M_{1} < M_{2}$). As shown in Fig.\ref{Nsub_pair}, subhaloes in high-mass haloes are more abundant than low-mass haloes in pair sample with respect to the control sample.

\section{The peak mass function of subhaloes}
\label{sec:Mpeak}
\begin{figure}
\centering
\includegraphics[width=\columnwidth]{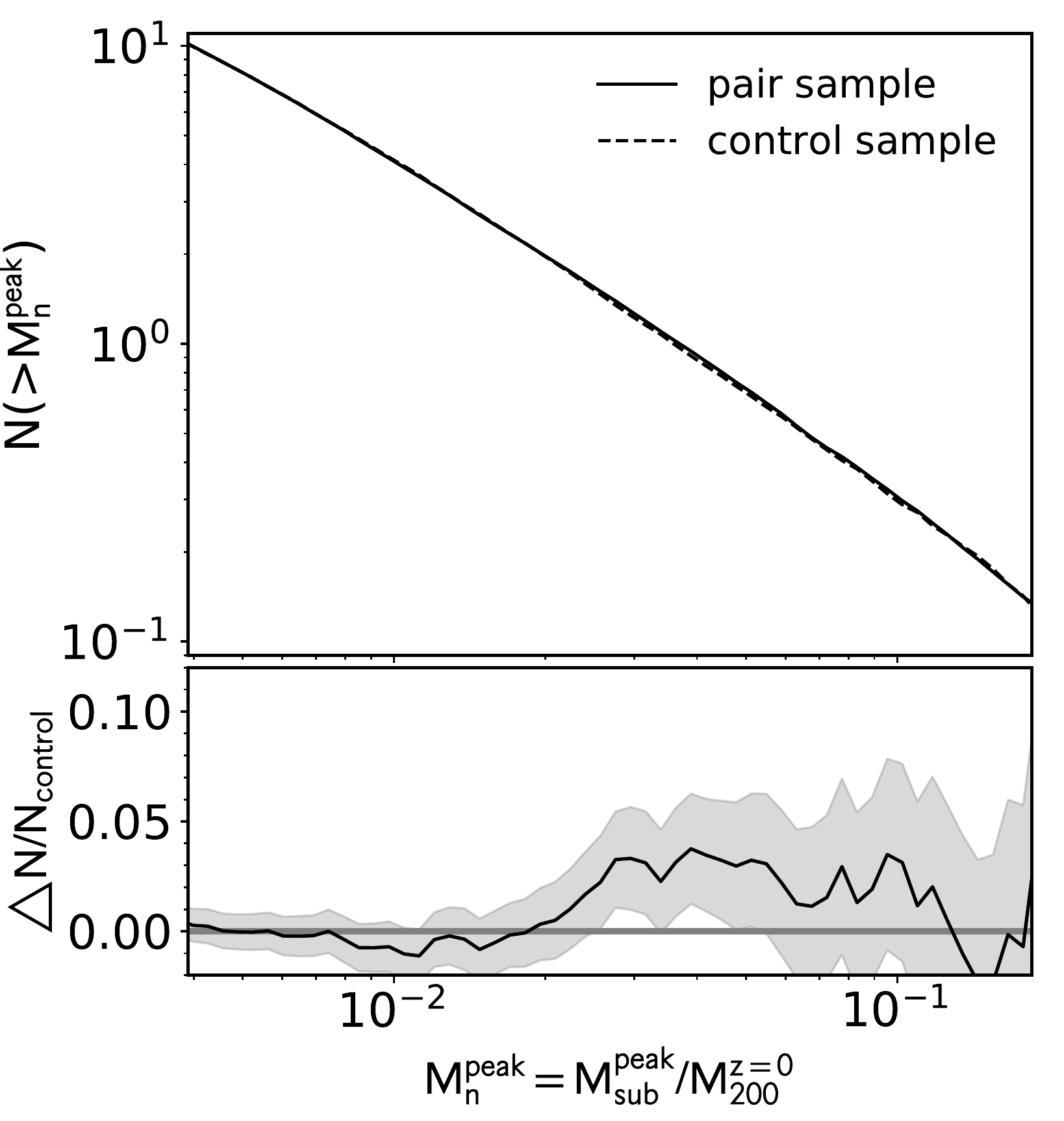}
\caption{\textit{Top panel}: the scaled peak subhalo mass function, $M^{\rm peak}_{\rm sub}/M_{\rm 200}^{\rm z=0}$, for type-I+II satellites in the pair (solid lines) and in the control sample (dashed lines) in the MS-II. \textit{Bottom panel}: relative difference of the peak subhalo abundance between the two samples. The shadow region represents 1$\sigma$ scatter calculated from the \textit{bootstrap} resampling method.}
\label{Npeak_sat} 
\end{figure}

Fig.\ref{Npeak_sat} shows the scaled peak mass function of subhaloes between the control and pair samples in the MS-II. We track back both the Type-I and Type-II satellites prior to their infall when their subhalo reached their peak mass, $M^{\rm peak}_{\rm sub}$. The subhaloes in pair sample is $\sim$2\% more abundant than those in the control sample, the result is consistent with the satellite abundance when Type-II satellites are included, as shown in the black lines in \reffig{Nminus_sat}.

\end{document}